\documentclass[10pt]{article}

\usepackage{amsmath,amsbsy,amssymb,latexsym,amsthm,dsfont}

\newtheorem{Theorem}{Theorem}
\newtheorem{Definition}{Definition}
\newtheorem{Remark}{Remark}

\newtheorem{example}{Example}

\newenvironment{keywords}{\begin{center}
\begin{minipage}[c]{12cm} {\bf Keywords:}} {\end{minipage}
\end{center}}
\newenvironment{msc}{\begin{center}
\begin{minipage}[c]{12cm} {\bf Mathematics Subject Classification:}} {\end{minipage}
\end{center}}

\begin{document}

\title{Generalized Euler-Lagrange equations
for variational problems with scale derivatives\thanks{Submitted on 03-Aug-2009; accepted for publication 16-March-2010;
in \emph{Letters in Mathematical Physics}.}}

\author{Ricardo Almeida\\
\texttt{ricardo.almeida@ua.pt}
\and
Delfim F. M. Torres\\
\texttt{delfim@ua.pt}}

\date{Department of Mathematics\\
University of Aveiro\\
3810-193 Aveiro, Portugal}

\maketitle


\begin{abstract}
We obtain several Euler-Lagrange equations for variational
functionals defined on a set of H\"{o}lder curves.
The cases when the Lagrangian contains multiple scale derivatives,
depends on a parameter, or contains higher-order scale derivatives
are considered.
\end{abstract}

\begin{msc}
49K05, 26B05, 39A12.
\end{msc}

\begin{keywords}
scale calculus, Euler-Lagrange equations,
higher-order scale derivatives, non-differentiability.
\end{keywords}


\section{Introduction}

In 1992 L.~Nottale introduced the theory of scale-relativity
without the hypothesis of space-time differentiability
\cite{Nottale:1992,Nottale:1999}. A rigorous mathematical
foundation to Nottale's scale-relativity theory was then given
by J.~Cresson in 2005 \cite{Cresson2,CD:Cresson:2006}.
Roughly speaking, the calculus of variations developed in \cite{Cresson2}
cover sets of non differentiable curves, by substituting the classical derivative
by a new complex operator, known as the \emph{scale derivative}.
Here we proceed with the theory started in \cite{Cresson2} and continued in \cite{Almeida,Cresson3},
by presenting several Euler-Lagrange equations in the class of H\"{o}lderian curves
and generalizing the previous results. For the necessary terminology and motivation
to the study of such non-differentiable calculus of variations
we refer the reader to \cite{Almeida,Cresson1,Cresson2,Nottale:1992}.

The paper is organized as follows. In Section~\ref{sec2}
we review the necessary notions of scale calculus.
Our results are then given in Section~\ref{sec:MR}:
(i) in \S\ref{sec3} we prove the Euler-Lagrange equation for functionals
defined by multiple scale derivatives; (ii) in \S\ref{sec4} we deduce
the Euler-Lagrange equation with a dependence on a complex parameter $\xi$;
(iii) finally, in \S\ref{sec5} we characterize the extremals
of variational functionals containing higher-order scale derivatives.


\section{Preliminaries}
\label{sec2}

In what follows, function $f\in C^0$ and $\epsilon>0$ is a real number.
The $\epsilon$-left and $\epsilon$-right quantum derivatives are defined by the formulas
$$\triangle ^-_{\epsilon}f(x)=-\frac{f(x-\epsilon)-f(x)}{\epsilon} \quad \mbox{ and } \quad
 \triangle ^+_{\epsilon}f(x)=\frac{f(x+\epsilon)-f(x)}{\epsilon},$$
respectively. The $\epsilon$ scale derivative of $f$ at $x$ is given by
$$\Box_{\epsilon}f(x)=\frac12 (\triangle ^+_{\epsilon}f(x)+\triangle ^-_{\epsilon}f(x))-
i\frac12(\triangle ^+_{\epsilon}f(x)-\triangle ^-_{\epsilon}f(x)).$$

\begin{Theorem}[\cite{Cresson2}]
\label{intparts}
Let $f,g\in C^0$ and $\epsilon>0$. Then,
$$\Box_{\epsilon}(f \cdot g)=\Box_{\epsilon}f \cdot g + f \cdot
\Box_{\epsilon}g+i\frac{\epsilon}{2}(\Box_{\epsilon} f \Box_{\epsilon}g-\boxminus_{\epsilon}f
\Box_{\epsilon}g-\Box_{\epsilon}f\boxminus_{\epsilon}g
-\boxminus_{\epsilon}f\boxminus_{\epsilon}g),$$
where $\boxminus_{\epsilon} (*)$ is the complex conjugate of $\Box_{\epsilon}(*)$.
\end{Theorem}

The space of curves under consideration is the space of H\"{o}lderian curves
with H\"{o}lder exponent $\alpha$. More precisely, given a real number $\alpha\in(0,1)$
and a sufficiently small real parameter $\epsilon$, $0<\epsilon \ll 1$, we define
$$C^{\alpha}_{\epsilon}(a,b) = \{ y:[a-\epsilon,b+\epsilon]
\to \mathbb R \, | \, y \in H^{\alpha} \}.$$ In \cite{Cresson2}
variational functionals of the type $\Phi(y)=\int_a^b f(x,y(x),
\Box_{\epsilon}y(x))\, dx $ are studied in the class $y \in C^{\alpha}_{\epsilon}(a,b)$.
As variation curves, it is considered those of the following type:
let $\beta$ be a positive real satisfying condition
$$
\beta \geq (1-\alpha) \mathds{1} _{]0,1/2[} + \alpha \mathds{1} _{[1/2,1]} \, ,
$$
and $h \in  C^{\beta}_{\epsilon}(a,b)$ be such that $h(a)=0=h(b)$.
A curve of the form $y+h$ is called a variation of $y$.
One says that $\Phi$ is differentiable on $C^{\alpha}_{\epsilon}(a,b)$ if,
for all curves $y \in C^{\alpha}_{\epsilon}(a,b)$ and for all variations $y+h$,
$$\Phi(y+h)-\Phi(y)=F_{y}(h)+ R_{y}(h),$$
where $F_{y}$ is a linear operator on the space $C^{\beta}_{\epsilon}(a,b)$
and $R_{y}(h)=O(h^2)$. A curve $y$ is an extremal for $\Phi$ on
$C^{\beta}_{\epsilon}(a,b)$ if $[F_y(h)]_{\epsilon}=0$ for all
$\epsilon>0$ and all $h \in C^{\beta}_{\epsilon}(a,b)$.

\begin{Theorem}[\cite{Cresson2}]
\label{th:tbg}
The curve $y$ is an extremal for $\Phi$ on
$C^{\beta}_{\epsilon}(a,b)$ if and only if
$$\left[ \partial_2 f(x,y(x), \Box_{\epsilon}y(x))-\Box_{\epsilon}(\partial_3 f)(x,y(x),
\Box_{\epsilon}y(x)) \right]_\epsilon=0$$
for every $\epsilon>0$.
\end{Theorem}


\section{Main Results}
\label{sec:MR}

We obtain several generalizations of Theorem~\ref{th:tbg}:
Theorem~\ref{ELtheorem} of \S\ref{sec3} coincides with Theorem~\ref{th:tbg}
in the particular case $n = 1$; Theorem~\ref{thm:withP} of \S\ref{sec4} coincides
with Theorem~\ref{th:tbg} in the particular case when the Lagrangian $L$
does not depend on the parameter $\xi$; Theorem~\ref{thm:ho2} of \S\ref{sec5} coincides
with Theorem~\ref{th:tbg} in the particular case when the Lagrangian $L$
does not depend on $\Box^2_{\epsilon}y(x)$ and $\xi$.


\subsection{Euler-Lagrange equation for multiple scale derivatives}
\label{sec3}

Let $\epsilon:=(\epsilon_1,\ldots,\epsilon_n)\in\mathbb R^n$ be a vector such that
$0<\epsilon_k \ll 1$ for all $k\in \{1,\ldots,n\}$. For continuous functions $f$ let
$$\triangle ^\sigma_{\epsilon_k}f(x)=\sigma\frac{f(x+\sigma\epsilon_k)-f(x)}{\epsilon_k},\quad \sigma=\pm.$$
We define the $\epsilon_k$ scale derivative of $f$ at $x$ by the rule
$$\Box_{\epsilon_k}f(x)=\frac{1}{2}\left(\triangle ^+_{\epsilon_k}f(x)+\triangle ^-_{\epsilon_k}f(x)\right)
-i\frac{1}{2}\left(\triangle ^+_{\epsilon_k}f(x)-\triangle ^-_{\epsilon_k}f(x)\right).$$
Let $\alpha\in(0,1)$ be a real, $\epsilon_M:=\max \{ \epsilon_1,\ldots,\epsilon_n\}$, and
$$C^{\alpha}_{\epsilon_M}(a,b) = \{ y:[a-\epsilon_M,b+\epsilon_M] \to \mathbb R \, | \, y \in H^{\alpha} \}.$$
We consider the following type of functionals: $\Phi:C^{\alpha}_{\epsilon_M}(a,b)\to \mathbb C$ of form
\begin{equation}
\label{funct}
\Phi(y)=\int_a^b L(x,y(x), \Box_{\epsilon_1}y(x),\ldots,\Box_{\epsilon_n}y(x))\, dx \, ,
\end{equation}
where $L:\mathbb R^2\times\mathbb C^n \to \mathbb C$ is a
given $C^1$ function, called the Lagrangian. As in \cite{Cresson2},
we assume the Lagrangian $L$ to satisfy
$$\left\|  DL(x,y(x), \Box_{\epsilon_1}y(x),\ldots,\Box_{\epsilon_n}y(x)) \right\| \leq C$$
for all $x$ and $\epsilon_1>0,\ldots,\epsilon_n>0$, where $C$ is a positive constant,
$D$ denotes the differential, and $\| \cdot \|$ is a norm for matrices.

\begin{Definition}
We say that $\Phi$, as in (\ref{funct}), is differentiable on $C^{\alpha}_{\epsilon_M}(a,b)$
if for all curves $y \in C^{\alpha}_{\epsilon_M}(a,b)$ and for all variations
$y+h$, $h \in C^{\beta}_{\epsilon_M}(a,b)$, the equality
$$\Phi(y+h)-\Phi(y)=F_y(h)+ R_y(h)$$
holds with $F_y$ a linear operator on the space
$C^{\beta}_{\epsilon_M}(a,b)$ and $R_y(h)=O(h^2)$.
\end{Definition}

Let us determine the expression of $F_y(h)$. To simplify notation, let
$u:=(x,y, \Box_{\epsilon_1}y,\ldots,\Box_{\epsilon_n}y)$. Then,
\begin{equation*}
\begin{split}
\Phi(y &+ h) - \Phi(y) \\
&= \displaystyle \int_a^b \left[
L(x,y+h, \Box_{\epsilon_1}y+\Box_{\epsilon_1}h,\ldots,\Box_{\epsilon_n}y+\Box_{\epsilon_n}h)
-L(x,y, \Box_{\epsilon_1}y,\ldots,\Box_{\epsilon_n}y)\right] dx \\
& = \displaystyle\int_a^b \left[ \partial_2 L(u) \cdot h+ \partial_3 L(u) \cdot \Box_{\epsilon_1} h
+\cdots + \partial_{n+2} L(u) \cdot \Box_{\epsilon_n} h\right] dx +O(h^2).
\end{split}
\end{equation*}
Integrating by parts (see Theorem~\ref{intparts}), we obtain:
\begin{multline*}
F_y(h) = \displaystyle\int_a^b \Bigl[ \partial_2 L(u) \cdot h
+ \Box_{\epsilon_1} (\partial_3 L(u) \cdot h)-\Box_{\epsilon_1}(\partial_3 L(u))\cdot
h-i\frac{\epsilon_1}{2} \Sigma_{\epsilon_1}(\partial_3L(u),h)\\
+\ldots +  \Box_{\epsilon_n} (\partial_{n+2} L(u) \cdot h)-\Box_{\epsilon_n}(\partial_{n+2}
L(u))\cdot h-i\frac{\epsilon_n}{2} \Sigma_{\epsilon_n}(\partial_{n+2}L(u),h) \Bigr]dx,\\
\end{multline*}
where
$$\Sigma_{\epsilon_k}(p,q)= \Box_{\epsilon_k}p\Box_{\epsilon_k}q-\boxminus_{\epsilon_k}p\Box_{\epsilon_k}q
-\Box_{\epsilon_k}p\boxminus_{\epsilon_k}q-\boxminus_{\epsilon_k}p\boxminus_{\epsilon_k}q.$$
Therefore,
\begin{equation*}
\begin{split}
F_y(h)&=\displaystyle\int_a^b \left[ \partial_2 L(u) -\Box_{\epsilon_1}(\partial_3 L(u))
-\cdots - \Box_{\epsilon_n}(\partial_{n+2} L(u))\right] \cdot h \, dx\\
&\qquad +\displaystyle\int_a^b \left[ \Box_{\epsilon_1} (\partial_3 L(u) \cdot h)
+\ldots + \Box_{\epsilon_n} (\partial_{n+2} L(u) \cdot h)\right]dx\\
&\qquad - i  \int_a^b \left[\frac{\epsilon_1}{2} \Sigma_{\epsilon_1}(\partial_3L(u),h)
+\ldots +\frac{\epsilon_n}{2}\Sigma_{\epsilon_n}(\partial_{n+2}L(u),h)\right]\, dx.
\end{split}
\end{equation*}
We just proved the following result:
\begin{Theorem}
For all $\epsilon_1>0,\ldots,\epsilon_n>0$,
the functional $\Phi$ defined by (\ref{funct}) is differentiable, and
\begin{multline*}
F_y(h)=\displaystyle\int_a^b \left[ \partial_2 L(u) -\Box_{\epsilon_1}(\partial_3 L(u))-\ldots
- \Box_{\epsilon_n}(\partial_{n+2} L(u))\right] \cdot h \, dx\\
+\displaystyle\int_a^b \left[ \Box_{\epsilon_1} (\partial_3 L(u) \cdot h)
+\ldots + \Box_{\epsilon_n} (\partial_{n+2} L(u) \cdot h)\right]dx\\
- i  \int_a^b \left[\frac{\epsilon_1}{2} \Sigma_{\epsilon_1}(\partial_3L(u),h)
+\ldots +\frac{\epsilon_n}{2}\Sigma_{\epsilon_n}(\partial_{n+2}L(u),h)\right]\, dx.
\end{multline*}
\end{Theorem}

\begin{Definition}
Let $a(\epsilon)$, $\epsilon=(\epsilon_1,\ldots,\epsilon_n)$, be a real
or complex valued function. We denote by $[\cdot]_\epsilon$ the linear operator such that
$$a(\epsilon)-[a(\epsilon)]_\epsilon \xrightarrow[\epsilon
\rightarrow 0]{}0 \quad \mbox{and}\quad  [a(\epsilon)]_\epsilon=0
\mbox{ if } \lim_{\epsilon \rightarrow 0}a(\epsilon)=0.$$
\end{Definition}

\begin{Definition}
A curve $y$ is an extremal for $\Phi$ on $C^{\beta}_{\epsilon_M}(a,b)$
if $[F_y(h)]_\epsilon=0$ for all $\epsilon_1>0,\ldots,\epsilon_n>0$
and $h \in C^{\beta}_{\epsilon_M}(a,b)$.
\end{Definition}

We now prove an Euler-Lagrange type equation for functionals of type (\ref{funct}).

\begin{Theorem}
\label{ELtheorem}
A curve $y$ is an extremal for $\Phi$ defined by (\ref{funct}) on
$C^{\beta}_{\epsilon_M}(a,b)$ if and only if
\begin{equation}
\label{ELequation}
\left[\partial_2 L(u)-\sum_{k=1}^n \Box_{\epsilon_k}(\partial_{k+2} L) (u)\right]_\epsilon=0
\end{equation}
for all $\epsilon_1>0,\ldots,\epsilon_n>0$,
where $u=(x,y(x), \Box_{\epsilon_1}y(x),\ldots,\Box_{\epsilon_n}y(x))$.
\end{Theorem}

\begin{proof}
Let $k \in \{1,\ldots,n\}$. Then, for all $t \in [a,b]$
$$\sup_{x \in [t,t+\sigma\epsilon_k]}\left|\partial_{k+2}L(x,y(x),
\Box_{\epsilon_1}y(x),\ldots,\Box_{\epsilon_n}y(x))\right| \leq C_k \epsilon_m^{\alpha-1},$$
for some constant $C_k$, where $\epsilon_m:=\min\{ \epsilon_1,\ldots,\epsilon_n \}$.
Therefore, as $\epsilon$ goes to zero, the quantities
$$\int_a^b \Box_{\epsilon_k} (\partial_{k+2} L(u) \cdot h)\,dx \quad \mbox{and}
\quad \epsilon_k \int_a^b \Sigma_{\epsilon_k}(\partial_{k+2}L(u),h)\, dx$$
also go to zero (\textrm{cf.} \cite[Lemma 3.2]{Cresson2}).
Applying the bracket operator $[\cdot]_{\epsilon}$ to $F_y(h)$,
the formula (\ref{ELequation}) follows from the fundamental
lemma of the calculus of variations.
\end{proof}

In \cite{Almeida} we study the isoperimetric problem in the scale calculus context,
considering integral constraints containing scale derivatives. There,
with the help of an appropriate auxiliary function,
we prove a necessary condition for extremals.
We now extend the main result of \cite{Almeida}
to functionals containing multiple scale derivatives.

\begin{Definition}
Let $\Phi(y)=\int_a^b L(x,y(x), \Box_{\epsilon_1}y(x),\ldots,
\Box_{\epsilon_n}y(x))\, dx$ and $\Psi(y)=\int_a^b g(x,y(x),
\Box_{\epsilon_1}y(x),\ldots,\Box_{\epsilon_n}y(x))\, dx$
be two functionals on $C^{\alpha}_{\epsilon_M}(a,b)$.
A curve $y$ is called an extremal of $\Phi$ subject
to the constraint $\Psi(y)=c$, $c \in \mathbb C$,
if for all $m \in \mathbb N$ and all variations
$\hat y= y+\sum_{k=1}^m h_k$, where $(h_k)_{1\leq k \leq m}
\in C^{\beta}_{\epsilon_M}(a,b)$ are such that $\Psi(\hat y)=c$,
one has $[F_{y}(h_k)]_{\epsilon}=0$ for all $\epsilon_1>0,
\ldots,\epsilon_n>0$ and all $k\in\{1,\ldots,m\}$.
\end{Definition}

Using the techniques in the proof of Theorem~\ref{ELtheorem} and in
\cite[Theorem~4]{Almeida}, the following result can be easily obtained.

\begin{Theorem}
Suppose that $y \in C^{\alpha}_{\epsilon_M}(a,b)$ is an extremal for the functional
$\Phi$ on $C^{\beta}_{\epsilon_M}(a,b)$ subject to the constraint
$\Psi(y)=c$, $c \in \mathbb C$. If
\begin{enumerate}
\item{$y$ is not an extremal for $\Psi$;}
\item{both limits
$$\displaystyle \lim_{\epsilon\to0} \max_{x \in [a,b]} \left|
\partial_2 L-\Box_{\epsilon_1}(\partial_3 L) -\ldots
- \Box_{\epsilon_n}(\partial_{n+2} L) \right|$$
and
$$\displaystyle \lim_{\epsilon\to0} \max_{x \in [a,b]} \left|
\partial_2 g-\Box_{\epsilon_1}(\partial_3 g) -\ldots
-\Box_{\epsilon_n}(\partial_{n+2} g) \right|$$
are finite along $y$;}
\end{enumerate}
then there exists $\lambda \in \mathbb C$ such that
$$\left[ \partial_2 K-\Box_{\epsilon_1}(\partial_3 K) -\ldots
- \Box_{\epsilon_n}(\partial_{n+2} K)\right]_{\epsilon}=0$$
holds along the curve $y$, where $K=L-\lambda g$.
\end{Theorem}


\subsection{Dependence on a parameter}
\label{sec4}

We now characterize the extremals
in the case when the variational functional depends
on a complex parameter $\xi$.
We consider functionals of the form
\begin{equation}
\label{funct2}
\Phi(y,\xi)=\int_a^b L(x,y(x),
\Box_{\epsilon}y(x),\xi)\, dx,
\end{equation}
where $(y,\xi) \in C^{\alpha}_{\epsilon}(a,b)\times\mathbb C$.
We say that $\Phi$ is differentiable if
for all $(y,\xi) \in C^{\alpha}_{\epsilon}(a,b)\times\mathbb C$
and for all $(h,\delta) \in  C^{\beta}_{\epsilon}(a,b)\times\mathbb C$
one has
$$
\Phi(y+h,\xi+\delta)-\Phi(y,\xi)
=F_{(y,\xi)}(h,\delta) + R_{(y,\xi)}(h,\delta)\, ,
$$
where $F_{(y,\xi)}$ is a linear operator on the space
$C^{\beta}_{\epsilon}(a,b)\times\mathbb C$
and $R_{(y,\xi)}(h,\delta)=O(|(h,\delta)|^2)$.
With similar calculations as done before, we deduce:
\begin{equation*}
\begin{split}
\int_a^b \Bigl[ &L(x,y+h, \Box_{\epsilon}y+\Box_{\epsilon}h,\xi+\delta)
- L(x,y, \Box_{\epsilon}y,\xi)\Bigr]_{\epsilon}\, dx\\
&=\int_a^b \left[\partial_2L(u) \cdot h +
\partial_3 L(u) \cdot\Box_{\epsilon} h + \partial_4L(u) \cdot \delta\right]_{\epsilon} \, dx\\
&= \int_a^b \left[\partial_2L(u)- \Box_{\epsilon}( \partial_3 L(u))\right]_{\epsilon}\cdot
h \, dx + \int_a^b \left[ \partial_4L(u)\right]_{\epsilon} \cdot \delta \, dx,
\end{split}
\end{equation*}
where $u:=(x,y(x), \Box_{\epsilon}y(x),\xi)$. For $\delta=0$
we obtain the Euler-Lagrange equation $\left[\partial_2L(u)- \Box_{\epsilon}(
\partial_3 L(u))\right]_{\epsilon}=0$ for all $\epsilon>0$. For $h =0$
we get $\int_a^b \left[ \partial_4L(u)\right]_{\epsilon} \, dx=0$.
In summary, we have:

\begin{Theorem}
\label{thm:withP}
The pair $(y,\xi)$ is an extremal for $\Phi$ given by (\ref{funct2}), \textrm{i.e.},
$\left[F_{(y,\xi)}(h,\delta)\right]_{\epsilon}=0$,
if and only if
$$\left\{\begin{array}{l}
\left[\partial_2L(u)- \Box_{\epsilon}( \partial_3 L(u))\right]_{\epsilon}=0\\
\int_a^b \left[ \partial_4L(u)\right]_{\epsilon} \, dx=0\\
\end{array}\right.$$
for all $\epsilon>0$, where $u=(x,y(x), \Box_{\epsilon}y(x),\xi)$.
\end{Theorem}

\begin{example}
\label{ex:1}
Let $\Phi$ be given by the expression
$$
\Phi(y,\xi)
=\int_{-1}^1 (\Box_{\epsilon}y-\Box_{\epsilon}|x|)^2+(\xi x)^2\, dx \, .
$$
Then, $(y,\xi)=(|x|,0)$ is an extremal of $\Phi$:
$$
\left[\partial_2L(u)- \Box_{\epsilon}( \partial_3 L(u))\right]_{\epsilon}
=\left[ - \Box_{\epsilon}(2(\Box_{\epsilon}y-\Box_{\epsilon}|x|)) \right]_{\epsilon}=0
$$
and
$$
\int_{-1}^1 \left[ \partial_4L(u)\right]_{\epsilon} \, dx
=\int_{-1}^1 \left[ 2\xi x^2\right]_{\epsilon} \, dx=0.
$$
\end{example}

\begin{example}
Consider now the functional $\Phi(y,\xi)=\int_{-1}^1
(\xi \cdot \Box_{\epsilon}y-\Box_{\epsilon}|x|)^2\, dx$,
$\xi\in \mathbb{C}$. Similarly as in Example~\ref{ex:1},
it can be proved that $(y,\xi)=(|x|,1)$
is an extremal of $\Phi$. Observe that if we substitute
$y$ by $|x|$ and $\xi\in\mathbb{R}$ in $\Phi$, simple calculations show that
$\Phi(|x|,\xi)=2(\xi-1)^2(1-\epsilon)$. This function
has a global minimizer for $\xi = 1$.
\end{example}


\subsection{Higher-order Euler-Lagrange equation}
\label{sec5}

Let $\alpha\in(0,1)$, $\epsilon>0$,
$n\in \mathbb{N}$, and $f\in C^{n-1}[a,b]$
be a real valued function.

\begin{Definition}
For $k=1,\ldots,n$ let
$$\triangle ^{k,\sigma}_{\epsilon} f(x)=\sigma\frac{f^{(k-1)}(x
+\sigma\epsilon)- f^{(k-1)}(x)}{\epsilon},\quad \sigma=\pm.$$
We define the $k$th $\epsilon$-scale derivative of $f$ at $x$ by
$$
\Box^k_{\epsilon}f (x)=\frac{1}{2}\left(\triangle ^{k,+}_{\epsilon}f(x)
+\triangle ^{k,-}_{\epsilon}f(x)\right)
- i \frac{1}{2}\left(\triangle ^{k,+}_{\epsilon}f(x)-\triangle ^{k,-}_{\epsilon}f(x)\right).
$$
\end{Definition}
Consider variational functionals of the form
$$\Phi(y)=\int_a^b L(x,y(x),\Box^1_{\epsilon}y(x),\ldots,\Box^n_{\epsilon}y(x))\, dx,$$
for curves $y$ of class $C^{n-1}$ and $y,\Box^1_{\epsilon}y,\ldots,\Box^{n-1}_{\epsilon}y
\in C^{\alpha}_{\epsilon}(a,b)$.
Observe that, as $\epsilon\to0$, we obtain the standard
functional of the calculus of variations with higher-order derivatives:
$$\Phi(y)=\int_a^b L(x,y(x),y'(x),\ldots,y^{(n)}(x))\, dx.$$
We study the case $n=2$:
\begin{equation}
\label{funct3}
\Phi(y)=\int_a^b L(x,y(x),\Box^1_{\epsilon}y(x),\Box^2_{\epsilon}y(x)) dx \, .
\end{equation}
Results for the general case are easily proved by induction.
Let $h$ be a function of class $C^1$  such that $h,\Box^1_{\epsilon}h
\in C^{\beta}_{\epsilon}(a,b)$, $h(a)=0=h(b)$ and $h'(a)=0=h'(b)$.
Observe that $\Box^2_{\epsilon}y=\Box^1_{\epsilon}y'$
and $\Box^2_{\epsilon}h=\Box^1_{\epsilon}h'$. Thus,
by Theorem~\ref{intparts} and the standard integration by parts formula
(here we are assuming that $L$ and $y$ are at least of class $C^2$),
\begin{equation*}
\begin{split}
[\Phi(y+h) &-\Phi(y)]_{\epsilon}\\
&= \displaystyle \int_a^b  \left[L(x,y+h,\Box^1_{\epsilon}y
+\Box^1_{\epsilon}h,\Box^2_{\epsilon}y+\Box^2_{\epsilon}h)
-L(x,y,\Box^1_{\epsilon}y,\Box^2_{\epsilon}y)\right]_{\epsilon}\, dx\\
&= \displaystyle  \int_a^b  \left[\partial_2 L\cdot h+\partial_3 L\cdot
\Box^1_{\epsilon}h +\partial_4 L \cdot \Box^2_{\epsilon}h \right]_{\epsilon}\, dx \displaystyle +O(h^2)\\
&=  \displaystyle \int_a^b  \left[\partial_2 L\cdot h+\partial_3 L\cdot
\Box^1_{\epsilon}h +\partial_4 L \cdot \Box^1_{\epsilon}h' \right]_{\epsilon}\, dx+O(h^2)\\
&= \displaystyle  \int_a^b \left[\partial_2 L\cdot h- \Box^1_{\epsilon}(\partial_3 L)\cdot h
- \Box^1_{\epsilon}(\partial_4 L) \cdot h' \right]_{\epsilon}\, dx+O(h^2)\\
&=  \displaystyle \int_a^b  \left[\partial_2 L- \Box^1_{\epsilon}(\partial_3 L)
+ ( \Box^1_{\epsilon}(\partial_4 L))'  \right]_{\epsilon}\cdot h\, dx+O(h^2) \, .\\
\end{split}
\end{equation*}
We just deduced the Euler-Lagrange equation for \eqref{funct3}:

\begin{Theorem}
Let $L$ be a Lagrangian of class $C^2$, and $\Phi$ as in (\ref{funct3})
be defined on the class $C^2$ of curves such that
$y,\Box^1_{\epsilon}y \in C^{\alpha}_{\epsilon}(a,b)$.
Function $y$ is an extremal for $\Phi$ if and only if
$$\left[\partial_2 L- \Box^1_{\epsilon}(\partial_3 L)
+ ( \Box^1_{\epsilon}(\partial_4 L))' \right]_{\epsilon}=0$$
for all $\epsilon>0$.
\end{Theorem}

\begin{Remark}
In contrast with the classical
theory of the calculus of variations for functionals
containing second-order derivatives,
where typically admissible
functions are of class $C^4[a,b]$, here it is enough
to work with $C^2$ curves.
\end{Remark}

We can easily include the case when the functional depends
on a complex parameter $\xi$, as was done in Section~\ref{sec4}:

\begin{Theorem}
\label{thm:ho2}
Let $\Phi$ be the functional defined by
$$\Phi(y,\xi)=\int_a^b L(x,y(x),\Box^1_{\epsilon}y(x),
\Box^2_{\epsilon}y(x),\xi)\, dx \, .$$ The pair $(y,\xi)$
is an extremal for $\Phi$ if and only if
$$\left\{\begin{array}{l}
\left[\partial_2 L- \Box^1_{\epsilon}(\partial_3 L)
+ ( \Box^1_{\epsilon}(\partial_4 L))' \right]_{\epsilon}=0\\
\int_a^b \left[ \partial_5L(u)\right]_{\epsilon} \, dx=0\\
\end{array}\right.$$
for all $\epsilon>0$,
where $u=(x,y(x),\Box^1_{\epsilon}y(x),\Box^2_{\epsilon}y(x),\xi)$.
\end{Theorem}


\section*{Acknowledgments}

Work supported by the {\it Centre for Research on Optimization and Control} (CEOC)
from the ``Funda\c{c}\~{a}o para a Ci\^{e}ncia e a Tecnologia'' (FCT),
cofinanced by the European Community Fund FEDER/POCI 2010.



\end{document}